\documentclass[%
 reprint,
showpacs,preprintnumbers,
 amsmath,amssymb,
 aps,superscriptaddress,
 groupeaddress,
 floatfix,
]{revtex4-1}

\usepackage{array}
\usepackage{hhline}
\usepackage{natbib}
\usepackage{color}
\usepackage{arydshln}
\usepackage{graphicx}
\usepackage{dcolumn}
\usepackage{bm}
\usepackage{textcomp}
\usepackage{xspace}
\usepackage{soul}

\newcolumntype{d}{D{+}{\,\pm\,}{3,3}}
\newcolumntype{x}[1]{>{\centering\arraybackslash\hspace{0pt}}p{#1}}

\newcommand\T{\rule{0pt}{2.3ex}}
\newcommand\B{\rule[-1.2ex]{0pt}{0pt}}

\begin{document}

\title{Constraining the Neutron Star Compactness: Extraction of the 
$^{23}$Al($p,\gamma$) Reaction Rate for the $rp$-Process}


\author{C.~Wolf}
\affiliation{Institute for Applied Physics, Goethe University, 60438 Frankfurt am Main, Germany}
\author{C.~Langer}
\altaffiliation{Corresponding author}
\email{langer@iap.uni-frankfurt.de}
\affiliation{Institute for Applied Physics, Goethe University, 60438 Frankfurt am Main, Germany}
\author{F.~Montes}
\author{J.~Pereira}
\affiliation{National Superconducting Cyclotron Laboratory, Michigan State University, East Lansing, MI 48824, USA}
\affiliation{JINA Center for the Evolution of the Elements, Michigan State University, East Lansing, MI 48824, USA}
\author{W.-J.~Ong}
\affiliation{National Superconducting Cyclotron Laboratory, Michigan State University, East Lansing, MI 48824, USA}
\affiliation{JINA Center for the Evolution of the Elements, Michigan State University, East Lansing, MI 48824, USA}
\affiliation{Department of Physics and Astronomy, Michigan State University, East Lansing, MI 48824, USA}
\author{T.~Poxon-Pearson}
\affiliation{National Superconducting Cyclotron Laboratory, Michigan State University, East Lansing, MI 48824, USA}
\affiliation{Department of Physics and Astronomy, Michigan State University, East Lansing, MI 48824, USA}
\author{S.~Ahn}
\affiliation{National Superconducting Cyclotron Laboratory, Michigan State University, East Lansing, MI 48824, USA}
\affiliation{JINA Center for the Evolution of the Elements, Michigan State University, East Lansing, MI 48824, USA}
\author{S.~Ayoub}
\affiliation{National Superconducting Cyclotron Laboratory, Michigan State University, East Lansing, MI 48824, USA}
\affiliation{Department of Physics and Astronomy, Michigan State University, East Lansing, MI 48824, USA}
\author{T.~Baumann}
\affiliation{National Superconducting Cyclotron Laboratory, Michigan State University, East Lansing, MI 48824, USA}
\author{D.~Bazin}
\affiliation{National Superconducting Cyclotron Laboratory, Michigan State University, East Lansing, MI 48824, USA}
\author{P.C.~Bender}
\affiliation{National Superconducting Cyclotron Laboratory, Michigan State University, East Lansing, MI 48824, USA}
\author{B.A.~Brown}
\author{J.~Browne}
\affiliation{National Superconducting Cyclotron Laboratory, Michigan State University, East Lansing, MI 48824, USA}
\affiliation{Department of Physics and Astronomy, Michigan State University, East Lansing, MI 48824, USA}
\author{H.~Crawford}
\affiliation{Nuclear Science Division, Lawrence Berkeley National Laboratory, Berkeley, California 94720, USA}
\author{R.H.~Cyburt}
\affiliation{National Superconducting Cyclotron Laboratory, Michigan State University, East Lansing, MI 48824, USA}
\affiliation{JINA Center for the Evolution of the Elements, Michigan State University, East Lansing, MI 48824, USA}
\author{E.~Deleeuw}
\author{B.~Elman}
\affiliation{National Superconducting Cyclotron Laboratory, Michigan State University, East Lansing, MI 48824, USA}
\affiliation{Department of Physics and Astronomy, Michigan State University, East Lansing, MI 48824, USA}
\author{S.~Fiebiger}
\affiliation{Institute for Applied Physics, Goethe University, 60438 Frankfurt am Main, Germany}
\author{A.~Gade}
\affiliation{National Superconducting Cyclotron Laboratory, Michigan State University, East Lansing, MI 48824, USA}
\affiliation{JINA Center for the Evolution of the Elements, Michigan State University, East Lansing, MI 48824, USA}
\affiliation{Department of Physics and Astronomy, Michigan State University, East Lansing, MI 48824, USA}
\author{P.~Gastis}
\affiliation{Central Michigan University, Mount Pleasant, MI 48859, USA}
\affiliation{JINA Center for the Evolution of the Elements, Michigan State University, East Lansing, MI 48824, USA}
\author{S.~Lipschutz}
\author{B.~Longfellow}
\affiliation{National Superconducting Cyclotron Laboratory, Michigan State University, East Lansing, MI 48824, USA}
\affiliation{Department of Physics and Astronomy, Michigan State University, East Lansing, MI 48824, USA}
\author{Z.~Meisel}
\affiliation{Institute of Nuclear \& Particle Physics, Department of Physics \& Astronomy, Ohio University, Athens, OH 45701, USA}
\affiliation{JINA Center for the Evolution of the Elements, Michigan State University, East Lansing, MI 48824, USA}
\author{F.M.~Nunes}
\affiliation{National Superconducting Cyclotron Laboratory, Michigan State University, East Lansing, MI 48824, USA}
\affiliation{Department of Physics and Astronomy, Michigan State University, East Lansing, MI 48824, USA}
\author{G.~Perdikakis}
\affiliation{Central Michigan University, Mount Pleasant, MI 48859, USA}
\affiliation{National Superconducting Cyclotron Laboratory, Michigan State University, East Lansing, MI 48824, USA}
\affiliation{JINA Center for the Evolution of the Elements, Michigan State University, East Lansing, MI 48824, USA}
\author{R.~Reifarth}
\affiliation{Institute for Applied Physics, Goethe University, 60438 Frankfurt am Main, Germany}
\author{W.A.~Richter}
\affiliation{Department of Physics, University of Stellenbosch, Matieland 7602, South Africa}
\affiliation{iThemba LABS, P.O. Box 722, Somerset West 7129, South Africa}
\author{H.~Schatz}
\affiliation{National Superconducting Cyclotron Laboratory, Michigan State University, East Lansing, MI 48824, USA}
\affiliation{JINA Center for the Evolution of the Elements, Michigan State University, East Lansing, MI 48824, USA}
\affiliation{Department of Physics and Astronomy, Michigan State University, East Lansing, MI 48824, USA}
\author{K.~Schmidt}
\altaffiliation[Current address: ]{TU Dresden, Zellescher Weg 19, 01069 Dresden}
\affiliation{National Superconducting Cyclotron Laboratory, Michigan State University, East Lansing, MI 48824, USA}
\affiliation{JINA Center for the Evolution of the Elements, Michigan State University, East Lansing, MI 48824, USA}
\author{J.~Schmitt}
\affiliation{National Superconducting Cyclotron Laboratory, Michigan State University, East Lansing, MI 48824, USA}
\affiliation{JINA Center for the Evolution of the Elements, Michigan State University, East Lansing, MI 48824, USA}
\author{C.~Sullivan}
\affiliation{National Superconducting Cyclotron Laboratory, Michigan State University, East Lansing, MI 48824, USA}
\affiliation{JINA Center for the Evolution of the Elements, Michigan State University, East Lansing, MI 48824, USA}
\affiliation{Department of Physics and Astronomy, Michigan State University, East Lansing, MI 48824, USA}
\author{R.~Titus}
\affiliation{National Superconducting Cyclotron Laboratory, Michigan State University, East Lansing, MI 48824, USA}
\affiliation{Department of Physics and Astronomy, Michigan State University, East Lansing, MI 48824, USA}
\author{D.~Weisshaar}
\affiliation{National Superconducting Cyclotron Laboratory, Michigan State University, East Lansing, MI 48824, USA}
\author{P.J.~Woods}
\affiliation{University of Edinburgh, Edinburgh EH9 3JZ, United Kingdom}
\author{J.C.~Zamora}
\author{R.G.T.~Zegers}
\affiliation{National Superconducting Cyclotron Laboratory, Michigan State University, East Lansing, MI 48824, USA}
\affiliation{JINA Center for the Evolution of the Elements, Michigan State University, East Lansing, MI 48824, USA}
\affiliation{Department of Physics and Astronomy, Michigan State University, East Lansing, MI 48824, USA}

\date{\today}

\begin{abstract}
The $^{23}$Al($p,\gamma$)$^{24}$Si reaction is among the most important reactions driving the energy generation in Type-I X-ray bursts. 
However, the present reaction-rate uncertainty limits constraints on neutron star properties that can be achieved with burst model-observation comparisons.
Here, we present a novel technique for constraining this important reaction by combining the GRETINA array with the neutron detector LENDA coupled to the S800 spectrograph at the National Superconducting Cyclotron Laboratory. The $^{23}$Al($d,n$) reaction was used to populate the astrophysically 
important states in $^{24}$Si. This enables a measurement in complete kinematics for extracting all relevant inputs necessary to calculate the reaction rate.
For the first time, a predicted close-lying doublet of a 2$_2^+$ and (4$_1^+$,0$_2^+$) state in $^{24}$Si was disentangled, finally resolving conflicting results from two previous measurements. Moreover, it was possible to extract spectroscopic factors using GRETINA and LENDA simultaneously. 
This new technique may be
used to constrain other important reaction rates for various astrophysical scenarios.
\end{abstract}


\maketitle

\textit{Introduction}.$-$ 
Type-I X-ray bursts (XRBs), thermonuclear explosions powered by hydrogen and helium burning on the surface of accreting neutron stars, provide unique insights into the nature of matter at near and above nuclear densities~\cite{1981ApJS...45..389W, 1998PhR...294..167S,2006NuPhA.777..601S,0954-3899-45-9-093001}. 
Advances in modeling XRBs have created an opportunity to constrain the properties of the system and the underlying neutron star, such as the mass, radius, and composition and rate of accreted material. 
These models are sensitive to the underlying nuclear physics inputs, in particular, the nuclear reactions involved in the thermonuclear runaway.\\
Systematic surveys have identified the $^{23}{\rm Al}(p,\gamma)^{24}{\rm Si}$ reaction rate as having one of the most significant impacts on the XRB light curve~\cite{2008ApJS..178..110P,2016ApJ...830...55C}. In principle,
this reaction can siphon material from the $^{22}\rm{Mg}$ waiting-point, which is already in $(p,\gamma)-(\gamma,p)$ equilibrium with $^{23}\rm{Al}$ in the early part of the rapid proton-capture $(rp)$-process~\cite{1994ApJ...432..326V}. However, the extent to which such a bypass is possible is highly uncertain owing to the current $^{23}{\rm Al}(p,\gamma)^{24}{\rm Si}$ reaction-rate uncertainty. Cyburt \textit{et al.}~\cite{2016ApJ...830...55C} found that a scale-down factor of 30, determined from the existing experimental uncertainty, with respect to the \textit{recommended} value from REACLIB~\cite{0067-0049-189-1-240}, is sufficient to essentially remove this bypass, resulting in a significant increase of the light-curve rise time and a decrease of its convexity~\cite{2018arXiv181207155M}. Even more importantly, it has been shown recently that such a reduction of the reaction rate has a strong impact on the inferred neutron star compactness~\cite{2018arXiv181207155M}. By comparing the observed light curve to simulations, it is possible to constrain the distance $d\xi^{1/2}_{b}$ and the surface gravitational redshift $1+z$~\cite{2012ApJ...749...69Z, 1982ApJ...256..637A,1538-4357-671-2-L141,0004-637X-860-2-147}, which, in turn, can be used to extract the mass-to-radius ratio $M_{NS}/R_{NS}$, since they are directly related~\cite{2016ApJ...819...46L}. This technique offers a complementary approach to the method described in e.g.~\cite{2010ApJ...722...33S, 2016ARA&A..54..401O} for sources where the Eddington limit is not reached.
Meisel \textit{et al.} investigated the sensitivity of the distance 
and the gravitational redshift 
of the \textit{textbook} GS 1826-24 XRB source to uncertainties in several important nuclear reaction rates. Interestingly, they found that scaling the $^{23}{\rm Al}(p,\gamma)^{24}{\rm Si}$ reaction rate down by a factor of 30~\cite{2016ApJ...830...55C}
results in a drastic reduction of the surface gravitational redshift. Therefore, it is critical to reduce the $^{23}{\rm Al}(p,\gamma)^{24}{\rm Si}$ rate uncertainty in order to investigate the neutron-star compactness from simulation-observation comparisons.  
\\
For the $\sim$0.4~GK temperature where the $^{22}{\rm Mg}$ waiting-point bypass may be possible, $^{23}$Al($p$,$\gamma$)$^{24}$Si is mostly governed by resonant capture from the 5/2$^+$ ground state of $^{23}$Al into the first proton-unbound $2^{+}_{2}$ state of $^{24}$Si, with moderate contributions from direct capture into the ground and first excited (bound) $2^{+}_{1}$ state. At present, the main source of uncertainty for this reaction rate is in the excitation energy of the resonant $2^{+}_{2}$ state $E(2^{+}_{2})$, which defines the resonance energy $E_r=E(2^{+}_{2})-Q$. Whereas the $Q$-value is known with 19 keV uncertainty \cite{2017ChPhC..41c0002H}, only two measurements of $E(2^{+}_{2})$ exist, with conflicting results. In particular, the 3441(10) keV and 3410(16) keV values measured by Schatz \textit{et al.}~\cite{1997PhRvL..79.3845S} and Yoneda \textit{et al.}~\cite{2006PhRvC..74b1303Y}, respectively, differ by more than one standard deviation (1-$\sigma$). Since $E_r=E(2^{+}_{2})-Q$ enters exponentially in the calculation of the rate, this conflict leads to drastically different outcomes. As for the direct component of the reaction, the only existing experimental study was performed by Banu \textit{et al.}~\cite{2012PhRvC..86a5806B} to extract an asymptotic normalization coefficient (ANC) for the 
ground state of $^{24}$Si using one-proton breakup reactions. An experimental extraction of the spectroscopic factor of the $2^{+}_{1}$ state is still necessary.
\\
In the present work we aimed at measuring all the relevant sources of uncertainty in the $^{23}$Al($p,\gamma$)$^{24}$Si reaction rate using a novel technique based on the complete kinematics measurement of the $^{23}$Al($d$,$n$)$^{24}$Si transfer reaction. As the ($d,n$) reaction preferentially populates the states that might be of astrophysical interest at intermediate beam energies of 30~-~80~MeV/u t\cite{2014PhRvL.113c2502L, 2016EPJA...52....6K, 2017PhLB..769..549K}, it can be used to measure excitation energies, partial population cross sections from which spectroscopic factors C$^2$S can be inferred, spins and $\gamma$-ray decay cascades. Two recent ($d,n$) experiments at intermediate beam energies have been successfully employed to that effect studying the key novae reactions $^{26}$Al($p,\gamma$)$^{27}$Si and $^{30}$P($p,\gamma$)$^{31}$S by measuring the $\gamma$-rays emitted in the de-excitation of populated states and their corresponding "angle-integrated" cross section inferring the C$^2$S \cite{2016EPJA...52....6K,2017PhLB..769..549K}.
In this Letter, we demonstrate a novel ($d,n$)-based technique that takes the final step by measuring the full kinematics: $\gamma$-rays, projectile-like heavy recoils, and low-energy ejectile neutrons. This new approach opens up the possibility for complete measurements in a single experiment for astrophysically important reaction rates far from stability.

\textit{Experiment}.$-$ The $^{23}$Al($d,n$)$^{24}$Si experiment was carried out at the National Superconducting Cyclotron Laboratory at Michigan State University. To produce the 
secondary $^{23}$Al beam, a stable $^{24}$Mg beam with an intensity of 60 pnA at an energy of 170 MeV/u impinged on a 1904~mg/cm$^{2}$ thick $^9$Be production target located at the entrance of 
the A1900 fragment separator \cite{morrissey2003commissioning}. Since multi-nucleon transfer and pickup reactions produced a mix of several isotopes \cite{2014PhRvL.113c2502L, 2018EPJWC.16501055W}, the resulting beam was subsequently purified 
using the standard $\textrm{B}\rho-\Delta \textrm{E}-\textrm{B}\rho$ separation technique (with a 1050~mg/cm$^{2}$ Al wedge) in the A1900 fragment separator. Only isotopes within a 2\% 
momentum spread acceptance were transmitted. A beam with an average intensity of roughly 8$\times10^3$ pps of $^{23}$Al with a purity of 13\% (other isotonic admixtures from mainly $^{22}$Mg and $^{21}$Na) was 
subsequently delivered to the target position of the S800 magnetic spectrograph \cite{2003NIMPB.204..629B} where the ($d,n$) reaction took place. 
The high-resolution GRETINA $\gamma$-ray tracking device~\cite{2013NIMPA.709...44P, 2017NIMPA.847..187W} with eight detector modules mounted in one of the hemispheres was used to detect in-flight $^{24}$Si $\gamma$-ray de-excitations, while the other hemisphere was removed to provide space for the neutron detection.
The thickness of 110(5) mg/cm$^2$ for the CD$_2$ reaction target and the $^{23}$Al beam energy of 48 MeV/u were intended to maximize the
reaction yield. 
A pure C target of 78(4)~mg/cm$^2$ was also used in dedicated background runs; the extracted data taken with the C target were scaled and subtracted from the data taken with the CD$_2$ target in the analysis. The detection efficiency of GRETINA in singles mode was calibrated using standard sources and is about 4\% at an energy of 1800~keV.
\begin{figure}[b]
\begin{center}
\includegraphics[width=0.5\textwidth, keepaspectratio=true]{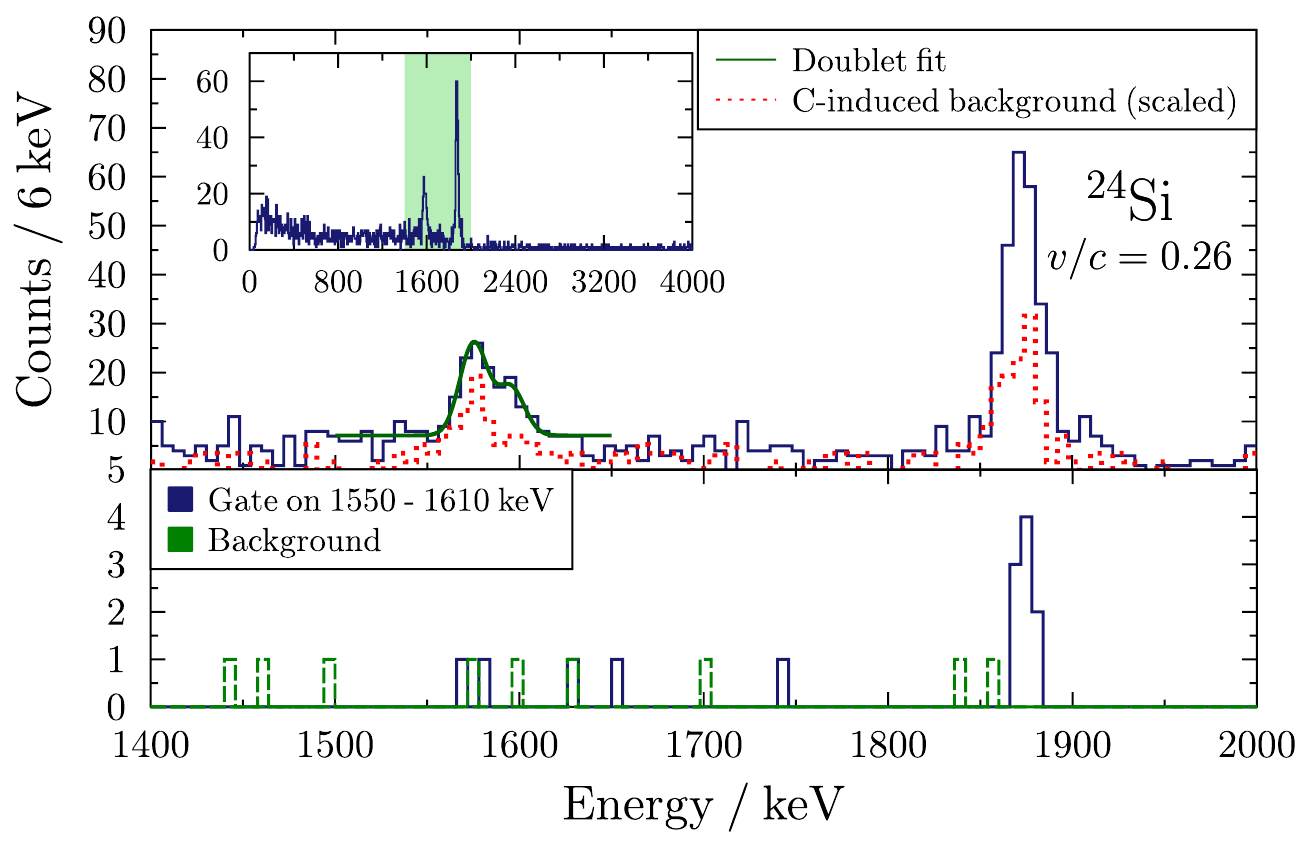}
\caption{\label{fig_gspectrum}Top: Doppler-corrected $\gamma$-ray spectrum of decays in $^{24}$Si in the $1400-2000$~keV range measured with GRETINA when gating on $^{24}$Si ions in the S800 focal plane (inset: full spectrum). The dotted histogram displays the scaled background induced by transfer-like
 reactions on carbon atoms in the target (see text). 
Bottom: The $\gamma$-ray coincidence spectrum gated on the $E_\gamma$ region between 1550 to 1610~keV. The coincidence with the $\gamma$-rays at 1874~keV
can be clearly identified.}
\end{center}
\end{figure}
\\
For detection of the neutrons, the low-energy neutron detector array (LENDA) \cite{2012NIMPA.686..117P, 2016NIMPA.815....1L} was used. In this experiment, LENDA consisted of 24 plastic-scintillation detector bars ($300\times45 \times 25$~mm each) which were installed in two vertical rows located at 1~m and 1.1~m distance from the S800 target covering scattering angles between 115~$\le \theta_{\text{lab}} \le$~175 degrees with no gaps between the bars.
The neutron time-of-flight (TOF) (used to determine the neutron kinetic energy) was derived from the difference between the neutron timing signal from LENDA and the recoil time signal from the S800 focal plane scintillator. The TOF resolution, as determined from the prompt-coincidence events in which a photon was emitted, was 744(36)~ps (FWHM) \cite{2018EPJWC.16501055W}.
\\
The $^{24}$Si recoil was separated and identified using its energy-loss, tracked positions, and time-of-flight measured with the S800 focal-plane detectors \cite{2003NIMPB.204..629B}. Because of the beam velocity of $\beta=v/c\approx0.26$, a Doppler correction of the deposited $\gamma$-ray energy in GRETINA was performed event-by-event after tracking the angle and the velocity of the residual nucleus $^{24}$Si through the S800 magnetic spectrograph with an ion-dynamical calculation using the COSY code \cite{2006NIMPA.558..346M}. Excitation energies (E$_x$) were deduced from the measurement of the $\gamma$-ray energies after Doppler correction. Together with the known reaction $Q$-value of 3293(20) keV \cite{2017ChPhC..41c0002H}, resonance energies $E_r=E_x-Q$ 
were obtained.\\
Due to the relatively thick CD$_2$ target combined with the 2\% momentum uncertainty of the incoming $^{23}$Al beam, the neutron energy resolution was $\sim$1.1 MeV which does not allow to resolve individual states \cite{2018EPJWC.16501055W}. Nevertheless, the detection of neutrons allowed for extraction of valuable spectroscopic information.

\textit{Results}.$-$ 
Figure~\ref{fig_gspectrum} shows the Doppler-corrected $\gamma$-ray spectrum measured with GRETINA when gating on incoming $^{23}$Al and on $^{24}$Si in the outgoing channel. A nearest-neighbor add-back algorithm \cite{2017NIMPA.847..187W} was applied to 
improve the peak-to-background 
ratio. 
Optimum parameters for the Doppler reconstruction were determined using known energy transitions from nuclei transmitted to the S800 focal plane in this experiment, like $^{18}$Ne (1489~keV and 1887~keV), $^{23}$Mg (1599~keV), $^{20}$Ne (1633~keV). Still, the Doppler correction remains the major source of uncertainty when determining the excitation energies.
The contribution measured with a pure carbon target when gating on $^{24}$Si recoils in the S800 focal plane  for background subtraction is also shown in Fig.~\ref{fig_gspectrum} (dashed line).
This contribution stemming from different proton-stripping reactions on carbon, like $(^{12}$C,$^{11}$B), was scaled to match the target thickness and the integrated beam on the CD$_2$ target (see also \cite{2016EPJA...52....6K,2017PhLB..769..549K}).
\\In order to extract the low-lying level scheme of $^{24}$Si (T$_z=+2$), experimental information from the isobaric T$_z=-2$ mirror nucleus $^{24}$Ne was compared to our observations.
The strongest transition at E$_\gamma=1874(3)$~keV can be assigned to the de-excitation of the bound $2_1^+$ state, similarly observed in previous experiments at energies of E$_\gamma=1879(11)$~keV \cite{1997PhRvL..79.3845S} and E$_\gamma=1860(10)$~keV \cite{2006PhRvC..74b1303Y}, respectively. 
This assignment is supported by the first excited 2$^+$ state in the mirror nucleus $^{24}$Ne at an energy of E$_x=1982(1)$~keV resulting in a Coulomb shift of 108~keV between the isobaric partners.\\
A broad structure is observed at an energy of about 1590~keV. Based on this peak width ($\sim$1.9\% FWHM) compared to the peak widths measured for other nuclei in the same experiment (1.1~-~1.4\% FWHM), it can be identified as two partially overlapping $\gamma$-ray transitions. 
Using a fit consisting of two Gaussian functions on top of a background contribution, two $\gamma$-ray energies of 1575(3)~keV and 1597(5)~keV were extracted. Both of these $\gamma$-rays are in coincidence with the 2$_1^+\rightarrow$~g.s. transition ($E_\gamma=1874(3)$~keV) as shown in the bottom panel of Fig.~\ref{fig_gspectrum}. Consequently, this yields excitation energies of E$_x=3449(5)$~keV and E$_x=3471(6)$~keV, respectively.
\begin{table*}[t]
\begin{center}
\caption{\label{tab_one}
Extracted level energies ($E_x$) and corresponding decay $\gamma$-ray energies ($E_\gamma$) in $^{24}$Si measured in this experiment. The (tentative) assignments are indicated. Different $l$ and $j$ values are presented for each state. Calculated cross sections within the ADWA theory ($\sigma_\text{theo}$) are compared to the measured partial cross sections ($\sigma_\text{exp}$) in this experiment. Theoretical spectroscopic factors (C$^2$S$_\text{theo}$), proton width ($\Gamma_p$), the $\gamma$-width ($\Gamma_\gamma$) and the spectroscopic strength ($\omega\gamma$) were calculated using the UDSB-cdpn shell model with the adapted C$^2$S$_\text{exp}$. For determination of C$^2$S$_\text{exp}$ see text and Eq.~\ref{eqnc2s}. No final assignment of the 3471~keV state is possible within this work; possible assignments are listed. The extracted cross section for the ground state is 168(103)~$\mu$b; however, we state the upper limit. The total $(d,n)$ cross section is 563(67)~$\mu$b.}
\vspace{0.14cm} 
\begin{tabular}{cclccclclccc}
\hhline{============}
$E_x$\T\B&$E_\gamma$&$J_i^\pi\rightarrow J_f^\pi$&$l$&$j$&$\sigma_{\textrm{theo}}$&C$^2$S$_\textrm{theo}$&$\sigma_{\textrm{exp}}$&C$^2$S$_\textrm{exp}$&$\Gamma_p$&$\Gamma_\gamma$&$\omega\gamma$\\
(keV)&(keV)&&&&($\mu$b)&&($\mu$b)&&(eV)&(eV)&(eV)\\
\hhline{------------}
0\T\B&&&2&5/2&98&3.44&$\le$~271&$\le$~2.8&&&\\
\hhline{------------}
1874(3)\T\B&1874(3)&$2_1^+\rightarrow 0_\textrm{gs.}^+$&0&1/2&139&0.27&263(83)&0.6(2)&&&\\
\T\B&&&2&3/2&473&0.03&&0.07(2)&&&\\
\T\B&&&2&5/2&411&0.17&&0.4(1)&&&\\
\hhline{------------}
3449(5)\T\B&1575(3)&$(2_2^+)\rightarrow 2_1^+$&0&1/2&86&0.45&78(41)&0.7(4)&1.0$\times$10$^{-4}$&1.9$\times$10$^{-2}$&4.2$\times$10$^{-5}$\\
\T\B&&&2&3/2&402&0.001&&0.002(1)&&&\\
\T\B&&&2&5/2&349&0.176&&0.3(2)&&&\\
\hhline{------------}
3471(6)\T\B&1597(5)&$(4_1^+)\rightarrow 2_1^+$&2&3/2&722&0.016&54(30)&0.07(4)&7.0$\times$10$^{-7}$&5.0$\times$10$^{-4}$&5.2$\times$10$^{-7}$\\
\T\B&&&2&5/2&629&0.001&&0.004(3)&&&\\
\T\B&&$(0^+)\rightarrow 2_1^+$&2&5/2&69&0.24&54(30)&0.8(4)&6.2$\times$10$^{-5}$&1.6$\times$10$^{-3}$&5.0$\times$10$^{-6}$\\
\hhline{============}
\end{tabular}
\end{center}
\end{table*}
\\In the two previous experiments on $^{24}$Si, only one state was identified at excitation energies of E$_x=3441(10)$~keV \cite{1997PhRvL..79.3845S} and E$_x=3410(16)$~keV \cite{2006PhRvC..74b1303Y}, respectively. However, in the mirror as well as in the shell-model calculation of \cite{2006PhRvC..74b1303Y}, two states separated by just a few hundred keV were expected.
Taking into account the additional information from a detailed $^{24}$Si shell-model calculation using the USDB-cdpn interaction \cite{PhysRevC.74.034315} combined with the mirror, tentative spin-parities were assigned to the observed E$_x=3449(5)$~keV and E$_x=3471(6)$~keV states.
\\
The energy of the state at E$_x=3449(5)$~keV agrees very well with the already measured and assigned 2$_2^+$ state in \cite{1997PhRvL..79.3845S, 2010NuPhA.841..251I} at E$_x=3441(10)$~keV. Moreover, in the mirror as well as in the shell-model calculation, the 2$_2^+$ state is
lower in energy than the other states. Compared to the mirror, for which a roughly 10\% ground state decay branching is observed, no decay branching for this state is detected in this experiment for $^{24}$Si.
Based on these arguments, we tentatively assign the state at E$_x=3449(5)$~keV to be J$^\pi=2_2^+$. This results in a Coulomb shift of 419~keV compared to the isobaric 2$_2^+$ state in $^{24}$Ne at E$_x=3868$~keV. In principle, our new technique allows for assigning spin-parities based on the simultaneously measured angular distribution of the neutrons when requiring a $\gamma$-neutron coincidence. In this experiment, the statistics were too low to make use of this additional information.
\\
The next higher-lying state in the mirror and in the shell-model calculation is a 4$_1^+$ state at an energy of E$_x=3972$~keV and E$_x=3973$~keV, respectively. In the mirror, it decays exclusively through a $\gamma$-ray cascade to the 2$_1^+$ state (observed in this work at E$_x=1874(3)$~keV). 
The shell-model calculation, however, predicts a rather small spectroscopic factor for this 4$_1^+$ state, whereas the 
calculated single-particle cross section $\sigma_\text{theo}$ is large (see Tab.~\ref{tab_one}).
The next possible state in the mirror nucleus is the second 0$^+$ state at an energy of E$_x=4767$~keV, which also decays predominantly via 
the first 2$^+$ state followed by a transition to the ground state. If the state at E$_x=3471(6)$~keV in $^{24}$Si corresponds to this second 0$^+$, the Coulomb shift would be 
$\approx$~1300~keV, which is rather large: in \cite{2018PhRvC..97e4307L}
also downward shifts in excited states of $^{25}$Si
relative to those in the mirror nucleus $^{25}$Na
and from the USDB-cdpn calculation were observed. This was interpreted
as a Thomas-Ehrman shift for levels that are near and
above the proton-decay threshold (S$_p=3.29$~MeV),
and that have a relative large occupancy
of the $s_{1/2}$ proton orbital. For $^{24}$Si the $s_{1/2}$ proton orbital 
occupancies increase with excitation energy (from the USDB-cdpn calculation):
0.52 (g.s.), 0.72 (2$^+_1$), 0.97 (2$^+_2$), 0.49 (4$^+_1$) and 1.54 (0$^+_2$).
Thus, the Thomas-Ehrman shift is likely
responsible for the lowering of excited state energies.
It is therefore most likely that the experimental state at E$_x=3.471(6)$ MeV is the
4$^+_1$ shifted down from $^{24}$Ne by about 500 keV. 
If it were 0$^+_2$, the
shift from $^{24}$Ne would be the largest Thomas-Ehrman shift ever observed.
This shift is difficult to
calculate, since one should take into account the
two-proton decay channel
(S$_{(2p)}=3.43$~MeV). 
However, based on our data alone no clear conclusion can be drawn. We list both possibilities in Tab.~\ref{tab_one}.
\begin{figure}[t]
\begin{center}
\includegraphics[width=0.48\textwidth]{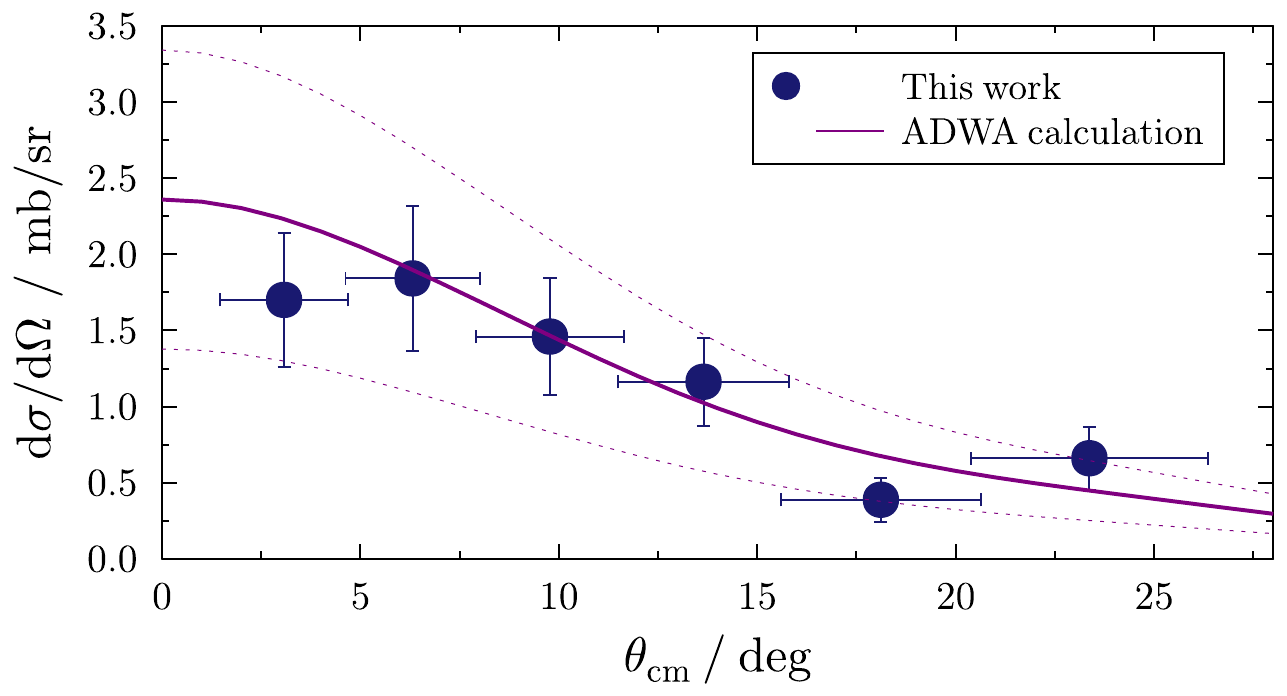}
\caption{\label{fig_nspectrum}Differential cross section in the center-of-mass system measured with the LENDA detector. Shown is the sum of all states
compared to a detailed ADWA calculation (purple line; see text). Error bands (dotted lines) are due to the uncertainty in the experimental spectroscopic factors.}
\end{center}
\end{figure}
\\Similar to the method described in \cite{2016EPJA...52....6K, 2017PhLB..769..549K}, the $\gamma$-ray transition intensities can be used to extract the integral and partial population cross sections after correcting for feeding. An integral $^{23}$Al($d,n$)$^{24}$Si cross section of 563(67)~$\mu$b at E$_{\textrm{beam}}=48$~MeV/u was extracted after subtracting the $^{24}$Si production due to carbon-induced reactions on the target using the pure C target runs and after correcting for the S800 momentum acceptance. 
Table~\ref{tab_one} summarizes the measured partial cross sections for states directly populated by the $^{23}$Al($d,n$)$^{24}$Si reaction. The partial cross section to the ground state was obtained by subtracting the sum of all excited states partial cross sections measured in GRETINA from the total cross 
section measured in the S800. 
\\
Partial single-particle cross sections for the $(d,n)$ transfer reactions were calculated using the FRESCO code \cite{thompson1988computer} in the adiabatic wave approximation (ADWA) \cite{1974NuPhA.235...56J} which explicitly incorporates deuteron breakup. The potentials and procedures used were the same as those implemented for \cite{2016EPJA...52....6K} (shown in Tab.~\ref{tab_one} as $\sigma_{\text{theo}}$).
Using the inferred ground-state partial cross section and the single-particle cross section, we obtain an upper limit of $\le$~2.8 for the ground-state spectroscopic factor. This is in excellent agreement with the value of 2.7(2) obtained in \cite{2012PhRvC..86a5806B}.
\\Theoretical spectroscopic factors,  C$^2$S$_\text{theo}$, were calculated with the shell model using the USDB-cdpn interaction. Table~\ref{tab_one} shows the shell-model theoretical spectroscopic factors and experimentally inferred C$^2$S$_\text{exp}$ that have been extracted for individual quantum numbers $i$ using the following relation:
\begin{equation}
\label{eqnc2s}
\text{C}^2\text{S}_{\text{exp}}^i=\frac{\text{C}^2\text{S}_{\text{theo}}^i\times \sigma_{\text{theo}}^i}{\sum_{i'}\left( \text{C}^2\text{S}_{\text{theo}}^{i'}\times\sigma_{\text{theo}}^{i'}\right)}\times\frac{\sigma_{\text{exp}}}{\sigma_{\text{theo}}^i}
\end{equation}
The neutron differential angular cross section in the center-of-mass frame measured with the LENDA detector is shown in Fig.~\ref{fig_nspectrum}. Due to the low yield of the reaction roughly 100 counts can be identified at energies between 6 and 16 MeV for the measured angles.  The relatively large thickness of the CD$_2$ reaction target, the low statistics, and the momentum spread of 2\% of the incoming beam
make it impossible to distinguish individual states in the neutron spectra by requiring a $\gamma$-neutron coincidence. Nevertheless, the total differential cross section in the center-of-mass frame can be used to independently verify the derived partial cross sections weighted by the experimentally extracted C$^2$S values from GRETINA (see Tab.~\ref{tab_one}). For each state and every single $l$-transfer, a specific angular distribution was calculated with the FRESCO code and weighted by our derived C$^2$S values. Eventually, all single contributions were added, which is shown as the solid line in Fig.~\ref{fig_nspectrum}. The dotted lines 
represent the uncertainty given by the uncertainty in the experimentally derived C$^2$S values. Using this method, a remarkable agreement is achieved (see Fig.~\ref{fig_nspectrum}). This confirms the possibility to perform these studies in complete kinematics and to extract all required information for the astrophysical reaction rate within one experiment (except for the equally important reaction $Q$-value).
\begin{figure}[t]
\begin{center}
\includegraphics[width=0.48\textwidth]{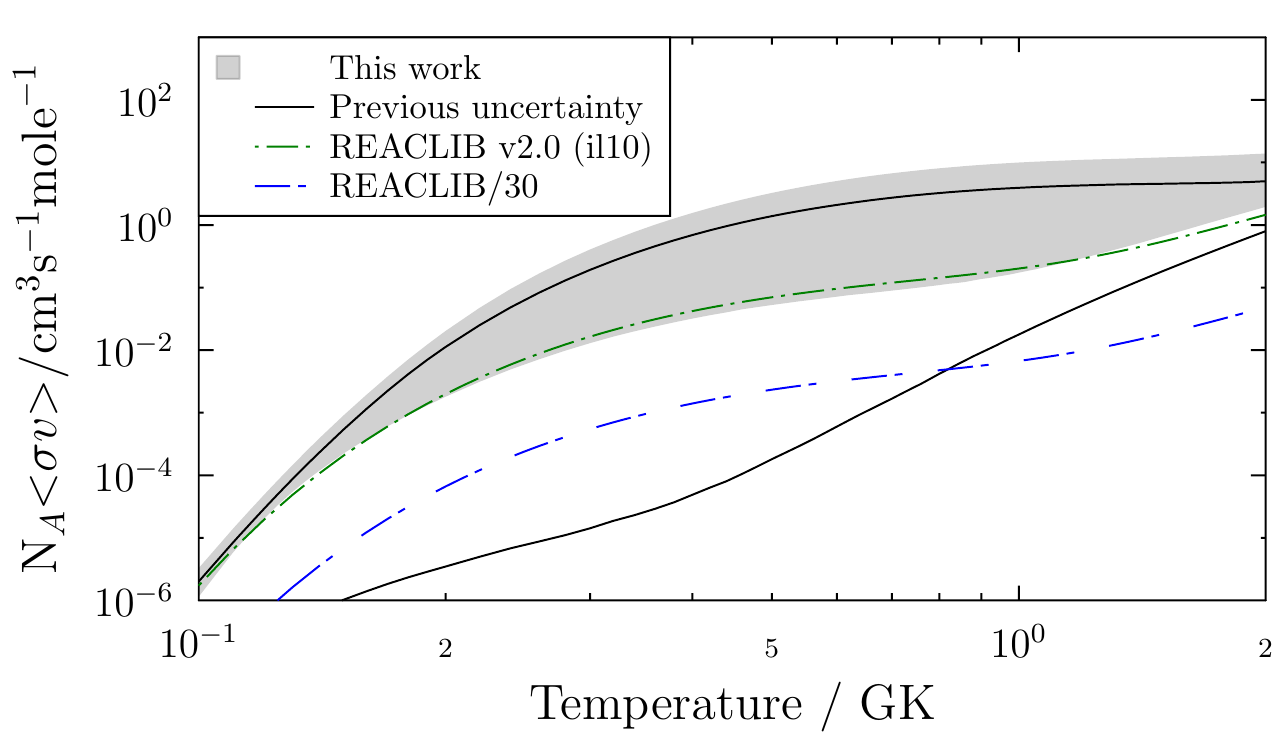}
\caption{\label{fig_rate}
The 1-$\sigma$ uncertainty band of the current rate (black lines) due to the conflicting results of \cite{1997PhRvL..79.3845S} and \cite{2006PhRvC..74b1303Y}. Also shown is the 1-$\sigma$ rate uncertainty calculated within this work (gray band). The calculations include the $Q$-value uncertainty of 20~keV.
Moreover, the recommended REACLIB (green) and REACLIB/30 (blue) rates are shown.}
\end{center}
\end{figure}

\textit{Astrophysical results}.$-$
Using the newly derived spectroscopic factors and the highly constrained excitation energies of the resonant states (with the J$^\pi=4^+_1$ assignment for the state at E$_x=3471(6)$~keV), a new reaction rate was calculated. The narrow resonance approximation \cite{2007nps..book.....I} was used with the excitation energies and the resonance strengths as input for each state above the proton separation energy (see Tab.~\ref{tab_one}). The direct-capture component was adapted from the most recent 
extraction in \cite{2012PhRvC..86a5806B} with a value of S(E$_0)=3.0812\times 10^{-3}$~MeV b.
The main uncertainty contribution stems from the resonance energy of the $2^+_2$ state which is due to the $Q$-value uncertainty of 19~keV.
In Fig.~\ref{fig_rate}, the upper and the lower
1-$\sigma$ error band of the new rate is shown (gray band) including all uncertainties in combination with the $Q$-value uncertainty. A Monte-Carlo based method with normally-distributed input parameters according to their experimentally given mean and variance was used.
Moreover, the 1-$\sigma$ rate uncertainty obtained by combining the experimental information available prior to the present work~\cite{1997PhRvL..79.3845S,2006PhRvC..74b1303Y} is included (solid black lines). As can be seen, the results presented here reduce the previous rate uncertainty by as much as 3-4 orders of magnitude in the temperature region of interest for XRBs. It is also worth emphasizing that, whereas the new rate is somewhat compatible with the values taken from the REACLIB database (green line), it clearly rules out the 30 scale-down factor determined by Cyburt \textit{et al.} on the basis of previously available experimental information~\cite{2016ApJ...830...55C} (blue line). 
With the results presented here, the extraction of a more precise gravitational red-shift in GS 1826-24 XRB will be possible using the method described in \cite{2018arXiv181207155M,0004-637X-860-2-147}. This, in turn, will result in an improved constraint of the neutron star compactness.
\\
\textit{Acknowledgments}.$-$
We thank the two anonymous referees for helpful comments. We would like to thank the operating staff at the NSCL for providing
an excellent beam. Support by Jeromy Tompkins and Ron Fox in merging the three data acquisition systems used in the present experiment is highly appreciated. Thanks to Lew Riley and his crew for the UCGretina simulation package. This work was supported by the US National Science Foundation (NSF) under Cooperative Agreement
No. PHY-1565546 (NSCL). GRETINA was funded by the DOE, Office of Science. Operation of the array at NSCL was supported by DOE under Grants No. DE-SC0014537
(NSCL) and No. DE-AC02-05CH11231 (LBNL). C.W. thanks JINA-CEE under NSF grant No.~PHY-1430152 for travel support and this research has received funding from the European Research Council under the European Unions's Seventh Framework Programme (FP/2007-2013) / ERC Grant Agreement n. 615126. C.L. thanks the Fokus A/B program of the Goethe University Frankfurt. Z.M. was supported by the U.S. Department of Energy under grant Nos.~DE-FG02-88ER40387 and DE-SC0019042. This work was supported in part by the National Nuclear Security Administration under the Stewardship Science Academic Alliances program through the U.S. DOE Cooperative Agreement No. DE-FG52-08NA2855 and under Award Number DE-NA0003180. B.A.B. received support from NSF grant PHY-1811855. W.R. received support from the National Research Foundation of South Africa Grant No. 105608.

\bibliographystyle{apsrev} 

\end{document}